# Evaluation of Static Analysis Tools for Finding Vulnerabilities in Java and C/C++ Source Code

Rahma Mahmood, Qusay H. Mahmoud
Department of Electrical, Computer & Software Engineering
University of Ontario Institute of Technology
Oshawa, ON, Canada
{rahma.mahmood1,qusay.mahmoud}@uoit.net

## Abstract

It is quite common for security testing to be delayed until after the software has been developed, but vulnerabilities may get noticed throughout the implementation phase and the earlier they are discovered, the easier and cheaper it will be to fix them. Software development processes such as the secure software development lifecycle incorporates security at every stage of the design and development process. Static code scanning tools find vulnerabilities in code by highlighting potential security flaws and offer examples on how to resolve them, and some may even modify the code to remove the susceptibility. This paper compares static analysis tools for Java and C/C++ source code, and explores their pros and cons.

## 1 Introduction

Ensuring security from the early stages of software development will save time, money and ensure a secure application environment. In fact, as shown in Fig. 1, the earlier a defect is discovered, the easier and cheaper it is to fix [2]. This requires testing source code as it is developed. With often thousands of lines of code to review for flaws, manual code review would serve to be an inefficient method of catching errors not caught by a debugger. Static code scanning tools may be used to find code vulnerabilities and discrepancies in programming code automatically. All scanners generally highlight potential security flaws. Most of these tools give examples on how to resolve the flaws while some may also modify the code to remove the susceptibility either automatically or each case is altered separately. Some static code analysis tools enable the user set rules themselves for locating susceptibilities or enforcing code standards [2].

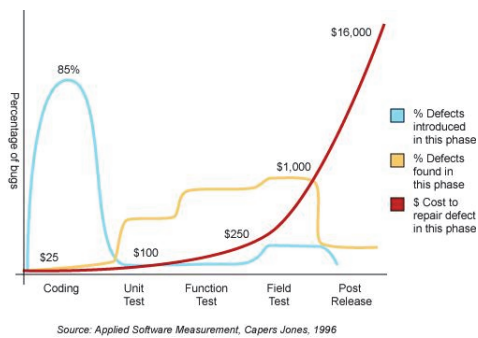

Fig. 1. Cost to repair defects during software development.

The rest of this paper is organized as follows. Section 2 discusses the related work. Java source code analyzers are evaluated in Section 3, and C/C++ code analyzers are discussed in Section 4. Conclusions are presented in Section 5.

## 2 Related Work

There have been many studies on static analysis tools focusing on their functionality or performance evaluation, such as those covered in [1, 3, 5, 7, 13]. Many of the studies focus on evaluation of a specific tool, for example in [7] the focus is on FindBugs. Other studies, such as previous work with the second author [5] cover several tools where they also evaluated tools using the Common Weakness Enumeration (CWE), which is a list of software security vulnerabilities found throughout the software development industry, and is a community-driven project maintained by MITRE.

## 3 Java Source Code Analyzers

The source code analyzers tested in this study scanned various approved and candidate source codes listed on the US Department of Homeland Security's Software Assurance Metrics and Tool Evaluation (SAMATE) Project website. These source codes contained common security vulnerabilities including: failure to sanitize directives in a web page (Cross-Site Scripting XSS); failure to sanitize data within SQL queries; unchecked error conditions; null pointer dereferences; unrestricted lock on critical sources; insufficient control of resource identifiers; OS command injection; hard coded passwords; left over debug code and TOC TOU (Time of Check to Time of Use).

### 3.1 FindBugs

FindBugs is a static code analysis tool that finds bugs in Java byte and source code. FindBugs essentially searches for potential problems by matching bytecode against a list of bug patterns [4]. Some of its strengths are that it successfully finds real defects in code and it has a low rate of detecting false bugs. Among its weaknesses is that it needs compiled code to work which can be quite troublesome to developers wanting to scan their code along the way during their development process [13].

*System Requirements*

FindBugs 3.0.0 requires minimum Java 7 for its runtime environment and supports Java 8. FindBugs may run on GNU/Linux, Windows, and MacOS X platforms [6].

*Effectiveness*

Out of the ten source code defects that FindBugs was tested to detect, FindBugs was only able to catch the NULL pointer dereferencing source code defect (see Fig. 2). However, in addition to this defect it brought to light some source code faults not mentioned by the source code authors. For example, one common error was "Reliance on default coding: Found a call to a method which will perform a byte to String (or String to byte) conversion, and will assume that the default platform encoding is suitable. This will cause the application behaviour to vary between platforms. Use an alternative API and specify a charset name or Charset object explicitly." [6]. This is a very important defect to catch because a developer may assume that the default UTF-8 setting is always used. However, a string's value may change in a different platform when the ASCII characters have different meanings. For example, instead of having an assigned value of A, on a different platform a string may have the value α. This is a simple example; the fact of the matter is that the purpose of the code and its performance may vary across different platforms if the character values change when an alternate API and charset name is not specified explicitly.

One weakness of FingBugs is that it requires compiled code in order to detect errors. If code cannot compile, FindBugs will not display any bugs. However, FindBugs displayed low false positives and found errors in code that were important to catch.

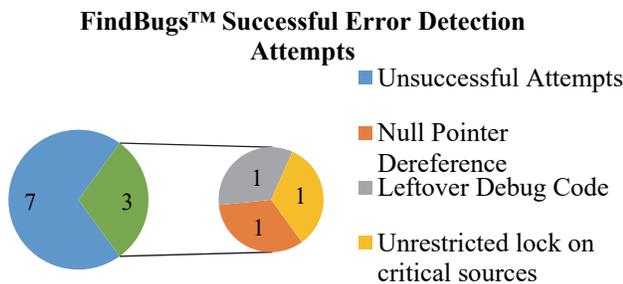

Fig. 2. Type and amount of successfully detected error types using FindBugs.

*Ease of Use*

FindBugs only requires the download of one JAR file and is easy to setup with the tutorial provided by OWASP. With step-by-step instructions and screenshots, installation is quick and easy. Furthermore, LAPSE+ scans all files in a project, making the process of scanning for source code errors less tedious since all vulnerabilities present in all source code files visible in one display. The Vulnerability Source View and Vulnerability Sink View provide basic descriptions for the detected errors. The Provenance Tracker View tracks the source vulnerability as its name states by displaying the propagation path the very source of the vulnerability, highlighting the problem in detail for the developer.

*Support*

FindBugs has documentation available detailing usage guidelines and types of vulnerabilities the plug-in may detect. However, not as many resources are available discussing the tool usage and potential problems as there are available for FindBugs and PMD on their websites.

## 3.2 PMD

PMD is a static analysis tool that looks for bugs such as empty statements, dead code, wasteful variable usage, over complicated expressions, and duplicated code (copied code means copied bugs) in Java source code. Some of its strengths are that it finds bad practices as well as real defects. Its primary weakness is that it is reported to be a slow duplicate code detector [13].

*System Requirements*

PMD is compatible with Mac OS X, Unix and Windows platforms. It is integrated with with JDeveloper, Eclipse, JEdit, JBuilder, BlueJ, CodeGuide, NetBeans/Sun Java Studio Enterprise/Creator, IntelliJ IDEA, TextPad, Maven, Ant, Gel, JCreator, and Emacs as a plug-in for each IDE [10]. The Eclipse plug-in was used for this study.

*Effectiveness*

PMD was used to scan ten source code flaws and it successfully caught potential problems that it claimed to catch such as finding empty try/catch statements and dead code (i.e. null pointer dereference) as shown in Fig. 3. In addition to finding these problems, PMD also identified complicated nested loops and other poor coding techniques such as uses of short variable names. PMD may help improve a developer's coding practices and style. Due to its large number of rulesets, PMD displays a great number of rule violations which may not always be important or risky violations at all and so a developer may choose to disregard warnings as he/she pleases. The large number of warnings may become troublesome with long source code files but PMD separates riskier warnings and errors by displaying warnings by category with high risk violations displayed at the top and styling ruleset violations towards the end. It is important to note that PMD does not claim to catch injection vulnerabilities and unsafe developing practices such as including hard coded passwords. For this reason, PMD should not be used solely as a security source code analyzer because it does not have the ability to identify high risk security holes such as resource injection.

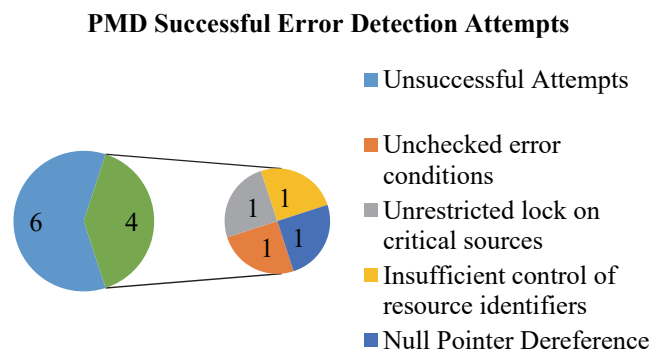

Fig. 3. Type and amount of successfully detected errors using PMD.

*Ease of Use*
PMD has a quick and easy setup with installation instructions available under its usage information in its online manual. Detailed instructions with screenshots are available for command line use and plug-in use both. The Eclipse plug-in is user friendly, clear, provides error details and separates errors based on violation type and risk level.

*Support*
PMD has detailed documentation available discussing previous version bugs, fixes, updates and frequently asked questions regarding the use of the plug-in. PMD is also a popular source code analyzer so it has a large community of users who may have solutions to issues or questions that may arise. The documentation on its own is detailed, thorough, and covers most, if not all, aspects of PMD plug-in usage.

### 3.3 LAPSE+

LAPSE is source code analyzer, developed by OWASP, which detects security vulnerabilities specifically for suspicious data injection in Java applications [8]. LAPSE+ only requires Eclipse IDE to analyze code and installation is simple and easy for its plug-in. The plug-in displays three views including: the *Vulnerability Sources View*, *Vulnerability Sinks View*, and the *Provenance Tracker View*. The first of these views points to lines in the source code being analyzed that can be a source of data injection. The *Vulnerability Sink View* specifies areas in code that are vulnerable to data insertion and manipulation. Finally, the *Provenance Tracker View* traces from a vulnerability sink to determine if a vulnerability source can be reached. If this tracker view traces back to a source then there is vulnerability in the code. LAPSE+ focuses specifically on finding areas of weakness for untrusted data injection.

*System Requirements*
LAPSE+ is used as a plugin for the Eclipse Java Development Environment, specifically functioning with Eclipse Helios and Java 1.6 or higher.

*Effectiveness*
Unlike FindBugs and PMD, LAPSE+ successfully detected OS command, SQL, and resource injection vulnerabilities after testing ten different source code weakness types (see Fig. 4). In doing so, this analyzer demonstrated its ability to fulfill its purpose. Although in the figure below it appears that LAPSE+ is not an effective tool since it only managed to detect 3/7 source code vulnerabilities, those three successful scans are the only types of vulnerabilities this tool is intended for. Further tests may be required to verify if LAPSE+ will always manage to detect injection type attacks. However, based on the results in this study, LAPSE+ was 100% successful in detecting the error types it claims to detect and therefore was found to be a highly effective tool.

*Ease of Use*
LAPSE+ only requires the download of one JAR file and is easy to setup with the tutorial provided by OWASP. With step-by-step instructions and screenshots, installation is quick and easy. Furthermore, LAPSE+ scans all files in a project, making the process of scanning for source code errors less tedious since all vulnerabilities present in all source code files visible in one display. The Vulnerability Source View and Vulnerability Sink View provide basic descriptions for the detected errors. The Provenance Tracker View tracks the source vulnerability as its name states by displaying the propagation path the very source of the vulnerability, highlighting the problem in detail for the developer.

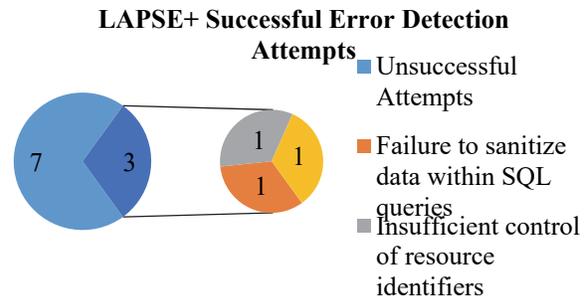

Fig. 4. Type and amount of successfully detected errors using LAPSE+.

*Support*
LAPSE+ has documentation available detailing usage guidelines and types of vulnerabilities the plug-in may detect. However, not as many resources are available discussing the tool usage and potential problems as there are available for FindBugs and PMD on their websites.

### 3.4 Yasca

Yasca is a flexible tool available for use with many source code languages including: Java, C/C++, PHP, COBOL, ASP, JavaScript, HTML, CSS, and Visual Basic [12]. Yasca detects security vulnerabilities and deviation from best practices in program source code. Overall, Yasca tries to improve quality of code and performance [9]. Yasca aggregates other external open source programs, such as FindBugs, PMD, JLint, JavaScript Lint, PHPLint, Cppcheck, ClamAV, RATS, and Pixy. In addition to those tools Yasca has custom scanning developed uniquely for itself as a standalone tool. Yasca runs using the command line tool.

*System Requirements*
Yasca requires minimum Java 1.5 to be installed for the plug-ins PMD, FindBugs, and Pixy to function. It has been tested to work successfully on Windows XP, Vista, Windows 7 as well as "a few flavours of Linux" [12]. Yasca was used on Windows 7 for the purposes of this study. The developer claims for Yasca to be extensible across multiple OS platforms however verification may be needed.

*Effectiveness*
Yasca was tested to be the most effective tool out of the four source code analyzers tested for Java. As shown in Fig. 5, Yasca successfully detected source code vulnerabilities for OS Command Injection, SQL injection, insufficient control of resource identifiers, unchecked error conditions and failure to sanitize data in a web page (Cross-Site Scripting). Yasca detected and displayed some of the flaws that PMD and FindBugs displayed but not all of them. Yasca also displayed less style warnings than PMD did and detected more vulnerabilities affecting the security of an application than PMD, FindBugs and LAPSE+ detected. While not as effective as PMD at catching poor coding style, it was evident that Yasca was more effective than FindBugs. Yasca detected the same injection type errors that LAPSE+ detected and more. Further tests and study would be needed to determine whether Yasca or LAPSE+ is more effective at detecting injection type errors. Overall, Yasca successfully found the most vulnerabilities tested for in this study.

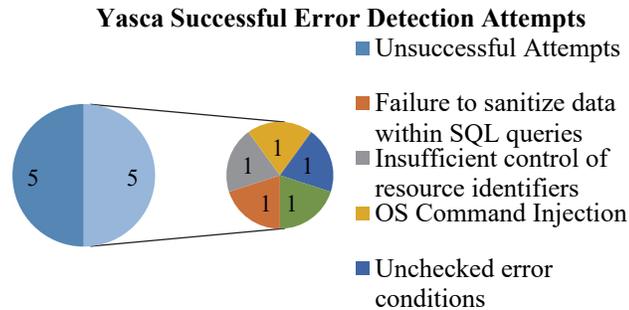

Fig. 5. Type and amount of successfully detected errors using Yasca.

*Ease of Use*

Unlike the other open source tools tested in this study, generates reports in HTML, CSV, XML, SQLite, and other formats [9]. Yasca requires command-line interface (CLI) knowledge but also comes with detailed instructions in readme files and documentation on its main web page on how to install and use Yasca and its plug-ins. Installation and setup was the most complex for this tool compared to the others because it required installing each external plug-in that it uses separately. Although Yasca does detect some source code vulnerabilities without the use of external plug-ins, it is certainly not as effective without them. All plug-ins were installed for this study and the additional effort is recommended.

*Support*

Yasca has documentation available detailing usage guidelines, system requirements, how the tool works, how it should be used and screenshots of examples. The readme files contained within the folders from the sourceforge download site are not as descriptive as the guide available online but should both be used if the use of additional external plug-ins is desired. The developer of this tool also provided an email on his website for questions and concerns regarding the use of his tool as well as to report any incompatibility issues.

Table 1. Java Static Analysis Tool Detected & Missed Errors

| Error Type | FindBugs | PMD | LAPSE+ | Yasca |
|---|---|---|---|---|
| CWE-259: Hard Coded Password | Missed | Missed | Missed | Missed |
| CWE-078: OS Command Injection | Missed | Missed | Found | Found |
| CWE-099: Insufficient Control of Resource Identifiers (Resource Injection) | Missed | Found | Found | Found |
| CWE-367: Time of Check to Time of Use | Missed | Missed | Missed | Missed |
| CWE-489: Leftover Debug Code | Found | Missed | Missed | Missed |
| CWE-476: NULL Pointer Dereference | Found | Found | Missed | Missed |
| CWE-391: Unchecked Error Condition | Missed | Found | Missed | Found |
| CWE-412: Unrestricted Lock on Critical Resource | Found | Found | Missed | Missed |
| CWE-079: Failure to Sanitize Directives in a Web Page (XSS) | Missed | Missed | Missed | Found |
| CWE-089: Failure to Sanitize Data within SQL Queries | Missed | Missed | Found | Found |
| Total # of Errors Detected | 3 | 4 | 3 | 5 |

### 3.5 Discussion

Each tool tested serves a particular function that is unique to itself. Although LAPSE+ was unable to detect other security holes, it is good at what it does and that is to detect path traversing and injection type errors. This tool may be used for this purpose alone as aid while developing code and testing for these weaknesses. PMD is an excellent tool for creating ones' own rule sets, detecting poor coding practices, NULL pointer dereferences and resource injection. Similarly, FindBugs was found to successfully detect NULL point referencing, critical resource lock vulnerabilities, as well as reliance on default coding. Yasca successfully detected the most vulnerabilities tested for in this study but also reported many false positives. If wanting to combine the functionality of FindBugs and PMD, Yasca aggregates the two tools in addition to other source code analyzers. Based on the developer's OS and IDE available, he/she may choose to use any, multiple or all of these tools to track each kind of vulnerability. Table 1 depicts the most number of errors may be detected if using both FindBugs and LAPSE+. I recommend using Yasca, LAPSE+ and FindBugs for testing if technical resources permit use of all tools.

## 4 C/C++ Source Code

The source code analyzers below were used to scan various approved source codes listed on the US Department of Homeland Security's Software Assurance Metrics and Tool Evaluation (SAMATE) Project website. These source codes contained common security vulnerabilities including: failure to sanitize directives in a web page (Cross-Site Scripting XSS), stack based buffer overflows, memory leaks, use of uninitialized variables, heap based buffer overflows, failure to sanitize data within SQL queries, unchecked error conditions, null pointer dereferences, insufficient control of resource identifiers, OS command injection, hard coded passwords, left over debug code, and buffer copy without checking size of input.

### 4.1 RATS

Rough Auditing Tool for Security (RATS) is a tool for scanning C, C++, Perl, PHP, and Python source codes. It also alerts the user by flagging frequent faults such as buffer overflows and TOCTOU (Time Of Check to Time Of Use) race conditions. RATS is a very useful tool but as stated in the name it only executes a rough analysis of the source code. This tool will not find every error in the source code and at times will also find "bugs" in a code which are not actually errors. RATS is a scanning tool that provides a list of probable concerning spots to target and it also defines the issues [11].

*System Requirements*
The RATS command line tool is only built and ready for installation for UNIX based platforms.

*Effectiveness*
RATS returned some successful results of finding security vulnerabilities but was not able to detect all known bugs (Fig. 6). RATS successfully found all buffer overflow errors, insufficient control of resource identifiers, OS command injection, and failure to constrain operations within the bounds of an allocated memory buffer. RATS was unsuccessful in detecting other errors. The figure below depicts the error catching results.

*Ease of Use*
Knowledge of Command Line Interface (CLI) use is required to use this tool. A GUI would make this user friendly and able to be used by non-technical individuals. Installation was relatively seamless after reading the README however there were several permission related issues when RATS tried to add files to usr/lib/bin which is defaulted to non-executable by most systems including OSX. Root is required to run the install. The README fails to mention this.

*Support*
The installation website should highlight installation process as users have to dig up README file to figure this out. Instructions are available but setup was not easy to do with instructions from README file alone.

In essence, RATS is a good tool to detect some types of buffer overflow cases and OS command injection in source code, however it is not easy to use without knowledge of CLI syntax. RATS has the ability to scan multiple source code languages including C, C++, Perl, PHP and Python which gives it a competitive advantage in terms of language flexibility over other open source static source code analyzers.

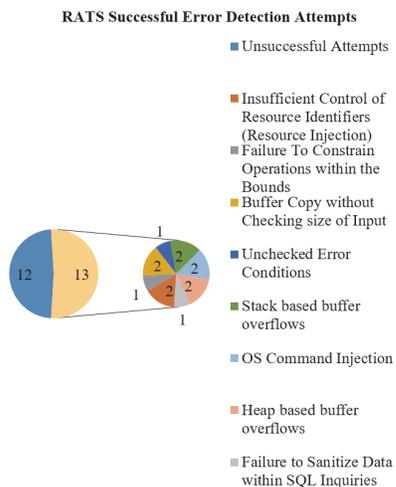

Fig. 6. Type and amount of successfully detected errors using RATS.

### 4.2 Flawfinder

Flawfinder is a code analysis tool that examines C/C++ code and reports possible flaws sorted by risk level [14]. Some advantages of Flawfinder are that it can handle internationalized programs (special calls like gettext()), it can also report column and line numbers of hits. Flawfinder is continually being updated and improved and has many resources available to help developers use this tool. Flawfinder falls short in its speed in comparison to RATS.

*System Requirements*
Flawfinder command line tool is only ready for installation and usage on Unix-like systems such as Linux, OpenBSD, or MacOS X.

*Effectiveness*
Flawfinder is officially compatible with CWE (Common Weakness Enumeration) and may detect many of SANS' Top 25 2011 list for most frequently occurring source code errors [14]. Among the CWE error cases that were used to run Flawfinder in this study, it detected CWE-078: OS Command Injection, CWE-119: Failure To Constrain Operations within the Bounds of an Allocated Memory Buffer and CWE-120: Buffer Copy without Checking Size of Input all successfully (Fig. 7). In addition to these weaknesses that Flawfinder claims to detect, it also detected stack based buffer overflows. Although Flawfinder was not able to detect every source code weakness used in this study, it delivered to its claims and detected the weaknesses in the figure below which is claims to detect. However, Flawfinder returned many false positives; even in events of a weakness not being present it occasionally displayed non-existent errors.

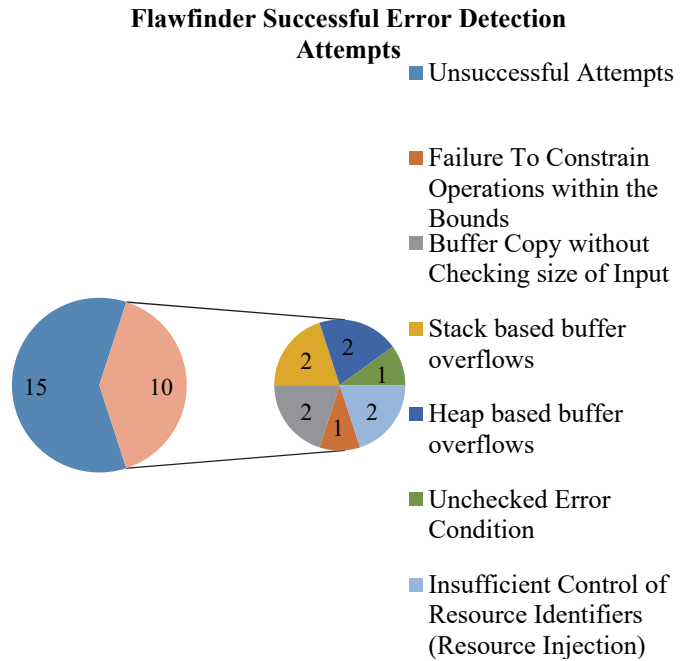

Fig. 7. Type and amount of successfully detected errors using Flawfinder.

*Ease of Use*

Flawfinder requires working knowledge of CLI on Unix-like systems but provides necessary command line inputs that are needed to use Flawfinder. Setup was not found to be as troublesome as the RATS tool was, and installation was quick with the aid of the Flawfinder manual available on the tool's main web page. Flawfinder's compatibility with CWE makes it particularly easier to understand the error types identified because lots of additional information on these error types is also available online through various sources.

*Support*

The Flawfinder web page, instruction manual and development are regularly updated. Moreover as mentioned above, additional information on the CWE error types that it detects are available online so that a developer may easily find why an error exists and what can be done to prevent similar errors in the future.

### 4.3 Yasca

Yasca is one of the tools that we have evaluated with Java as well as C++ code. Yasca quantitatively returned as many successful results as RATS. The successful results are displayed in Fig. 8. Yasca detected and displayed all of the same flaws that RATS displayed and it was evident that Yasca was more effective than Flawfinder. Yasca detected the same injection type errors that LAPSE+ detected and more. Overall, Yasca and RATS successfully found the most vulnerabilities tested for in this study.

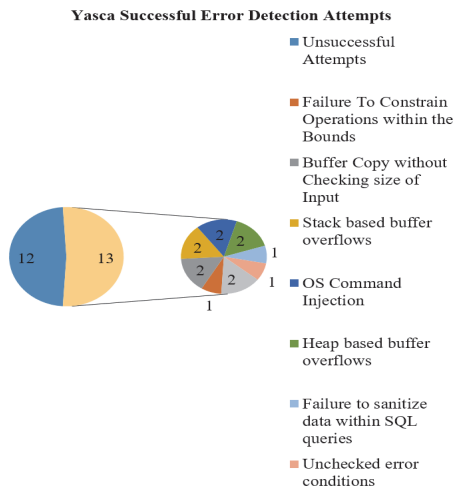

Fig. 8. Type and amount of successfully detected errors using Yasca.

### 4.4 Discussion

Flawfinder and RATS were found to be quite similar in the types of source code vulnerabilities that they detected. RATS was initially more difficult to install and set up compared to Flawfinder but it managed to detect more vulnerabilities than Flawfinder did. If combining the functionality of these tools is desirable, Yasca is also is available with a RATS plug-in along with many other source code analyzers and displays the same error information that the original tool would display in HTML format. Since Yasca is available with a RATS plug-in, it was able to detect the same vulnerabilities as RATS in addition to some more vulnerabilities that RATS would not have been able to detect. Based on the developer's OS and IDE available, he/she may choose to use any, multiple, or all of these tools to track each kind of vulnerability. Tables 2 and 3 depict the most number of errors may be detected if using either both RATS and Flawfinder or RATS and Yasca. We recommend using Yasca and Flawfinder if technical resources permit use of all tools because Yasca is more user friendly and generates reports with more information available than RATS.

Table 2. C++ Static Analysis Tool Detected & Missed Errors

| Error Type | RATS | FlawFinder | Yasca |
|---|---|---|---|
| CWE-079: Failure to Sanitize Web Directives (XSS) | Missed | Missed | Missed |
| CWE-121: Stack Based Buffer Overflow | Found | Found | Found |
| CWE-401: Memory Leak | Missed | Missed | Missed |
| CWE-457: Use of Uninitialized Variable | Missed | Missed | Missed |
| CWE-122: Heap Based Buffer Overflow | Found | Found | Found |
| CWE-089: Failure to Sanitize Data within SQL Queries | Missed | Missed | Missed |
| CWE-391: Unchecked Error Conditions | Missed | Found | Missed |
| CWE-476: NULL Pointer Dereference | Missed | Missed | Missed |
| CWE-099: Insufficient Control of Resource Identifiers | Found | Found | Found |
| CWE-078: OS Command Injection | Found | Missed | Found |
| CWE-259: Hardcoded Password | Missed | Missed | Missed |
| CWE-489: Leftover Debug Code | Missed | Missed | Missed |
| CWE-120: Buffer Copy without Checking Size of Input | Found | Found | Found |
| Total # of Errors Detected | 5 | 5 | 5 |

Table 3: C Static Analysis Tool Detected & Missed Errors

| Error Type | RATS | FlawFinder | Yasca |
|---|---|---|---|
| CWE-079: Failure to Sanitize Web Directives (XSS) | Missed | Missed | Missed |
| CWE-121: Stack Based Buffer Overflow | Found | Found | Found |
| CWE-122: Heap Based Buffer Overflow | Found | Found | Found |
| CWE-089: Failure to Sanitize Data within SQL Queries | Found | Missed | Found |
| CWE-391: Unchecked Error Conditions | Found | Missed | Found |
| CWE-476: NULL Pointer Dereference | Missed | Missed | Missed |
| CWE-099: Insufficient Control of Resource Identifiers | Found | Found | Found |
| CWE-078: OS Command Injection | Found | Missed | Found |
| CWE-259: Hardcoded Password | Missed | Missed | Missed |
| CWE-119: Failure To Constrain Operations within the Bounds of an Allocated Memory Buffer | Found | Found | Found |
| CWE-120: Buffer Copy without Checking Size of Input | Found | Found | Found |
| Total # of Errors Detected | 8 | 5 | 8 |

## 5 Conclusions

There are many tools available to analyze code, web applications, systems, and much more. However, a developer needs to be aware of common programming mistakes and how these flaws may compromise security. None of the tools analyzed in this study are perfect, and nor were they intended to be. There will often be false positives for "flaws" that are not actually harmful, and false negatives where flaws go undetected. Static analysis tools serve as a helpful hand to remind and detect common vulnerabilities that may accidentally go unnoticed. However, source code analyzers will not always point out and correct these issues. Developers should use these tools as an aid in the developmental process, but not depend on them. Creating safe and functional code is the developer's task and cannot be replaced by static code analysis tools. It is also important to note that a developer's choice of tool depends on the language being analyzed, the operating system, IDE, and types of vulnerabilities being sought. Indeed, there are numerous options available to detect security holes in software code during the developmental process. Adopting secure coding techniques and regularly using various methods of security vulnerability detection will surely reduce software security risks and improve efficiency for software developers.